# Elucidating the formation of structural defects in flax fibres through synchrotron X-ray phase-contrast microtomography

[a] Alain Bourmaud, [b] Lola Pinsard, [a] Elouan Guillou, [b]Emmanuel De Luycker, [b]Marina Fazzini, [c] Jonathan Perrin, [c] Timm Weitkamp, [b] Pierre Ouagne

[a]Univ. Bretagne Sud, UMR CNRS 6027, IRDL, Lorient, France
[b]Laboratoire Génie de Production, LGP, Université de Toulouse, INP-ENIT, Tarbes, France
[c]Synchrotron SOLEIL, Gif-sur-Yvette, France
*Corresponding author: alain.bourmaud@univ-ubs.fr

**Abstract**

The creation and ultrastructure of kink-bands in flax fibres are key issues for developing more and more performing biobased composite materials. Nevertheless, despite many hypotheses and structural characterization, the exact origin of kink-bands and the moment they appear remain unexplained. Here, by using cutting-edge techniques such as microtomography, a range of flax stems and fibres, from the green stem to stretched fibres, were morphologically explored. The study shows that all the extracted fibres, whether scutched, combed or stretched, contain significant amounts of kink-bands, which can be identified by the large pores they contain. On the other hand, at the scale of the green or retted stems, tomographic analysis does not reveal any kink-bands. These original observations suggest that the stress undergone by the plants during their growth is not sufficient, without major growth or lodging accidents, to generate these structural defects; the latter are only revealed after mechanical extraction of the fibres. Hypotheses regarding the kink-band appearance deformation levels are also given to complete the observations.

## 1. Introduction

Plant fibres, and particularly flax fibres, are receiving increasing interest as reinforcements for composite materials and, in Europe, an efficient industrial chain is already operating, whether it be for the extraction of fibres from plants, the manufacturing of reinforcements or the production of industrial parts.

Flax fibres are elementary cells made up of a primary wall and a secondary wall divided into three main layers S1, G and Gn (Rihouey et al., 2017). The G layer (also called S2) is the largest; it consists of about 80% crystalline cellulose and a matrix of non-cellulosic polymers (mainly pectins and hemicelluloses) (Roach et al., 2011). Within this layer, the microfibrils are oriented at about 5-7° to the fibre axis (Melelli et al., 2020); given its thickness (80-90% of that of the fibre), it is this layer that controls the mechanical properties of the fibre and therefore its potential to reinforce a composite material. However, this reinforcement potential can be affected by the presence of structural defects, called kink-bands. Within these defect zones, the structure of the fibre is modified, the orientation of the cellulose macro-fibrils changes strongly, locally up to 30-40° (Melelli et al., 2021a; Thygesen and Gierlinger, 2013). The cellulose presents discontinuities in terms of crystallinity, and pores can be noted (Aslan et al., 2010; Melelli et al., 2021a). All this contributes to the weakening of the fibre, and the kink-bands constitute preferential rupture zones (Sliseris et al., 2016) but also privileged entry points for water, micro-organisms or enzymes (Foulk et al., 2008; Hernandez-Estrada et al., 2016; Thygesen et al., 2011), inducing a more rapid alteration and degradation of the fibres due to their less homogeneous structure and the pores they contain. It has been shown, on very old flax fibres from the cultural heritage, that the kink-bands were more developed, suggesting a more marked ageing of the cells in these zones (Melelli et al., 2021b).

These defects within the fibres lead to increased fibre breakage, particularly when the fibres are subjected to high shear rates as may be the case when processing composites by extrusion or injection; these phenomena have been demonstrated by optical rheology (Le Duc et al., 2011) or morphological analysis of fibre lengths after processing (Gourier et al., 2017). Generally speaking, even if there is some controversy in the literature, it has been shown that the mechanical performance of elementary fibres is negatively impacted by the presence of kink-bands, in particular for tensile (Andersons et al., 2009; Baley, 2004; Grégoire et al., 2020; Zeng et al., 2015) or bending strength (Bos et al., 2002), which are more sensitive than modulus to the presence of defects.

Kink-bands can be identified using a large range of methods. Many authors use polarized light or electron microscopy (Bos et al., 2002; Gourier et al., 2017; Grégoire et al., 2020; Trivaudey et al., 2015), although the latter only highlights defects with a marked volume structure. It is also possible to access structural parameters of kink-bands (biochemical composition or micro-fibrillar angle, MFA) by Raman spectroscopy (Thygesen and Gierlinger, 2013) and recently developed techniques also allow access to the internal structure of kink-bands. This is the case of multi-photon microscopy, which has highlighted low-crystalline areas in the kink-bands, and atomic force microscopy (AFM), which, after fine preparation, makes it possible to visualize the pores at the core of these defects and also to characterize the stiffness of the cell walls inside the kink-bands (Melelli et al., 2021a). Thus, the available literature offers a good knowledge of the ultrastructure of these zones even if uncertainties may remain following the preparation of the samples. This is for example the case of internal pores which can be created by cutting or fracturing the samples before observation.

While the structure of kink-bands is well known and documented, the formation and development of kink-bands in fibres is not. Numerous hypotheses are present in the literature; for example, Thygesen et al. (Thygesen and Asgharipour, 2008) compared defect densities on carefully extracted hemp fibres from plants that had or had not been exposed to wind or drought during their growth and showed a greater quantity of kink-bands on plants exposed to wind or lacking water, which suggests that the bending of the stems would induce the formation of defects *in-planta*. This hypothesis is subject to the condition that the fibres are not altered during their extraction, which may be arguable given that the stems have not been retted. Given the low bucking performance of plant fibres, it has been shown by various authors that the mechanical extraction stages induce an increase in the density of defects in the fibres (Aslan et al., 2010; Grégoire et al., 2020), the scutching stage being generally considered as the most impactful; the refining of the fibres by carding also shows a constant increase in the number of defects present during the process. These different studies converge toward the fact that the stresses undergone by the fibres generate defects, especially during their extraction, but their development mechanisms in the stems are still unknown and the appearance of kink-bands at this stage is not proven. Recently, *in-planta* SEM observations confirmed this hypothesis; the obtained images, even if they are only local and cannot potentially highlight internal defects, do not evidence any kink-band before the mechanical extraction stages (Kozlova et al., 2022).

The present work has two major objectives:

- to study the presence of kink-bands in the fibres of stems at different stages of retting,
- to compare batches of scutched, combed and stretched fibres to detect potential differences in the structure and the density of these defects.

For this purpose, stems and fibres, from the same variety of flax, were studied by X-ray phase contrast microtomography using synchrotron radiation. Once the reconstruction parameters chosen and adapted to the acquisitions, the tomographic images of fibre sections and 3D representations were analyzed to characterize the presence or absence of defects and highlight the latter. Original movies were also designed to visualise the evolution of the internal structure of the fibre, and in particular the presence of pores in the kink-band zones.

## 2. Materials and methods

### 2.1. Flax plants and fibres

Seeds of industrial Flax Bolchoï variety were sowed at the end of March 2020 by Van Robaeys Frères company (Killem, France) in the north of France. Flax stems were pulled out in mid-July and then dew-retted in field for 7 weeks. Then, after full retting, fibres were successively extracted on an industrial scutching line and then industrially hackled. The hackled fibres were then processed at lab scale to produce a stretched strip. To complete the study, Bolchoï flax plants were cultivated in Tarbes (France) from March to June 2021, flax stems were pulled out when flowering, one day before the synchrotron investigations, to obtain green stems. In addition, a green stem was manually bent to artificially generate kink-bands. The sample range is summarized in Table 1.

Figure 1 summarizes the different cultivation, extraction and processing stages and illustrates the five specific materials we obtained for tomographic investigations.

### 2.2. SEM observations

Before tomographic investigation, fibres and stems were investigated by scanning electron microscopy (SEM) to assess the individualization rate, surface quality and also the presence of kink-bands. Prior to SEM investigations, samples were gold spun coated during 180 s (Edwards Scancoat Six device) and then analyzed with a Jeol JSM 6460LV scanning electron microscope (voltage 3kV).

**Table 1.** Sampling information

| Sampling | Variety | Cultivation year | Retting | Sample name |
|---|---|---|---|---|
| Stem | Bolchoï | 2021 | No | Green Stem |
| Stem | Bolchoï | 2021 | No | Bent Green Stem |
| Stem | Bolchoï | 2020 | 7 weeks | Retted Stem |
| Fibres | Bolchoï | 2021 | No | Green Fibres |
| Fibres | Bolchoï | 2020 | 7 weeks | Scutched Fibres |
| Fibres | Bolchoï | 2020 | 7 weeks | Hackled Fibres |
| Fibres | Bolchoï | 2020 | 7 weeks | Drawn Fibres |

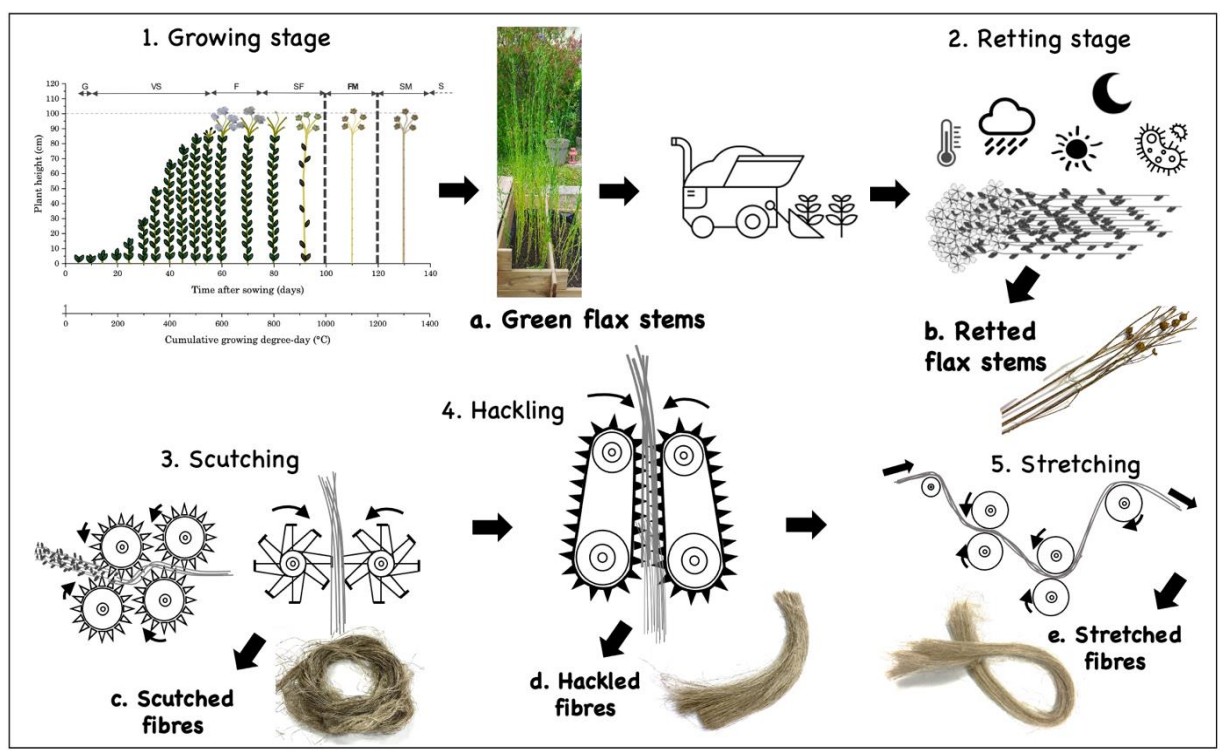

**Figure 1**: Cultivation, extraction and processing stages used in this study to obtain the five materials used: green flax stems (a), retted flax stems (b), scutched fibres (c), hackled fibres (d) and stretched fibres (e).

2.3. Synchrotron microtomography

X-ray phase contrast microtomography measurements were made at Synchrotron SOLEIL, the French national light source, on the ANATOMIX beamline (Weitkamp et al., 2017).

The fibres were mounted vertically free-standing on sample holders for a standard goniometer head (Huber Diffraktionstechnik, Rimsting, Germany). They were directly glued and scanned at a vertical position slightly above the glue, to avoid internal movement of the zone imaged during the scan.

The beamline was operated with an electron current of 500 mA in the SOLEIL storage ring and a gap of the U18 undulator X-ray source set to 10.7 mm. The polychromatic ("white") X-ray beam was filtered by a CVD diamond window of 0.6 mm thickness, resulting in a photon energy spectrum on the sample with a central value around 12 keV. The microtomography detector was a standard indirect design with a $Lu_3Al_5O_{12}$:Ce scintillator (Crytur, Turnov, Czech Republic) and a CMOS-based camera (Hamamatsu ORCA Flash 4.0 V2) with 2048 × 2048 pixels of size 6.5 µm, coupled via microscope optics using an objective 20×/NA 0.28 (Mitutoyo, Kawasaki, Japan). This resulted in a field of view of the detector of 0.65 × 0.65 mm² and an effective pixel size of 325 nm. The detector was placed 10 mm from the sample (distance scintillator–rotation axis). Each tomography scan contained 2000 equidistant projection angles over a range of 180°, with an exposure time of 100 ms per projection radiograph. The scans were made on-the-fly, i.e., the sample kept rotating during image acquisition. Volumes were reconstructed using the open-source software PyHST2 (ESRF, Grenoble, France) (Mirone et al., 2014). Applying a Paganin filter (Paganin et al., 2002) with a kernel length of 8 µm, the final resolution was estimated close to 1 µm.

2.4 3D Image analysis

The 3D visualization and analysis software Avizo 2021.1 of Thermo Fisher Scientific group was used for the image processing of the datasets, segmentation and analysis procedures. First, a Non-Local Means Filter was used to denoise image from the tomographic reconstructions. Then, segmentation process was conducted to visualize the objects in the images by defining a threshold intensity value. Due to the different absorption contrasts between the background and the fibre cell walls ~~tissues~~, the pores could be isolated from the rest of the fibre internal layers and the objects were highlighted with different colours.

2.5 Videos

A simultaneous displacement along the axis of the fibres in longitudinal and transverse views is proposed in a video format for: drawn fibres (video1), green stem (video2), bent green stem (video3), retted stem (video4) and extracted fibres from the green stem (video5). Both longitudinal view and transverse tomograms are extracted using FIJI software before being synchronized in Adobe Premier Pro software. In video1 and 5, free of defects areas are highlighted in green and kink-band regions in red. The coloured cursor in the top panel indicates the position of the slice shown in the bottom panel of the videos.

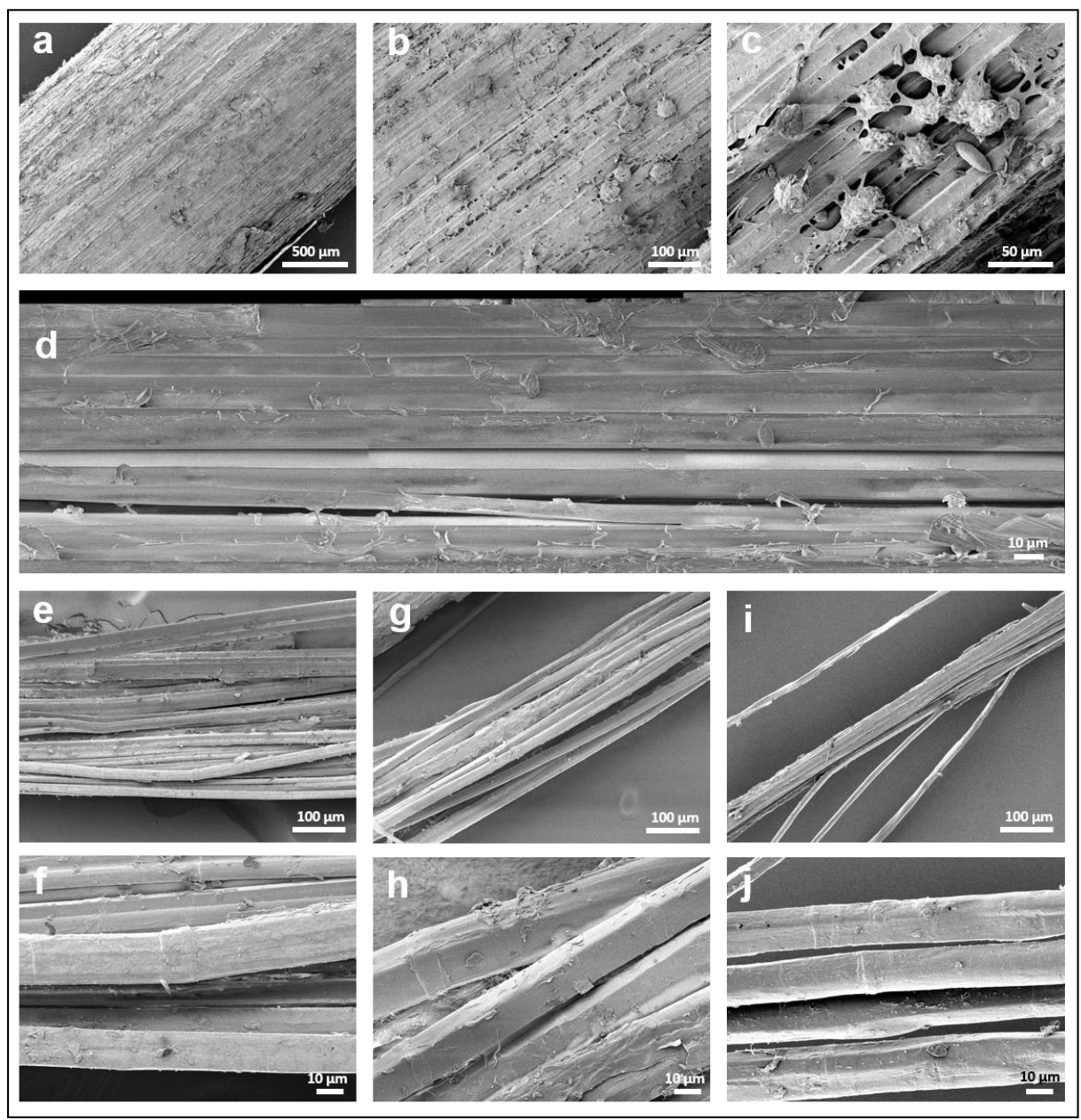

**Figure 2.** SEM observations of flax stems and fibres. Surface of a green stem (a), surface of a retted stem (b) with visible destructuration and traces of fungi with a specific zoom in (c). Longitudinal observation of fibres, visible in a highly damaged area of a retted stem (d). Finally, observation at two different magnifications of scutched (e, f), hackled (g, h) and stretched (I, j) fibres.

## 3. Results

### 3.1. Overall observations of fibres and stems

Figure 2 shows SEM observations of the different batches of stems and fibres. Panels a to d show views of the surface of the green (Fig. 2a) and retted (Fig. 2 b-d) stems. The surface of the green stem is very cohesive and does not appear damaged. On the other hand, the retted stems show degradation, which can reveal areas of fibre, as well as the presence of micro-organisms and fungi (Fig. 2c). Observations were made in areas where fibres were visible due to stem decohesion and cortical parenchymal damage. They allowed longitudinal reconstructions of the fibre zones to be made (Fig. 2d); with these observations, longitudinal morphology of a few fibres is followed over several hundred micrometers. The fibres visualized in this way do not show any obvious defects or kink-bands. However, we can note that these are only a few fibres which are in no way representative of all the fibre bundles present in the stem.

Scutched, hackled and drawn fibres were also observed by scanning electron microscopy (Fig. 2e-j). Two scales of magnification (x200 and x1000) were used. The effect of the successive mechanical treatments is clearly visible with a progressive decrease in the size of the fibre bundles; still very cohesive after the scutching stage, these bundles contain only a few fibres after drawing (Fig. 2i). Moreover, at each stage it is possible to observe the presence of kink-bands. Whatever the stage of transformation considered, this is not more pronounced and does not seem to be abundant either. These observations are linked to several hypotheses in the literature, Kozlova et al. (Kozlova et al., 2022) as well as Hendricks et al. (Hendrickx, 2019) who studied the influence of the processing history and hypothesized that kink-bands were indeed mainly created during the mechanical processing stages; our SEM observations carried out on fibre visible at the surface of stems highly corroborate the conclusions of Kozlova et al. (Kozlova et al., 2022) and the absence of visible kink-bands *in-planta* through SEM observations.

### 3.2. Evidencing the kink-bands through tomographic observations in an elementary flax fibre

One of the main advantages of tomography is to visualise the inner structure of the fibre such as the presence of pores or defects. The feasibility of tomographic investigations on single flax fibres has already been demonstrated by Abbey et al. (Abbey et al., 2010) but without enough resolution to explore fine details. Figure 3 shows tomographic slice images of an elementary stretched fibre along a longitudinal plane (Fig. 3a, h) and transverse planes (Fig. 3 b-g). Dark areas inside the flax fibres in the images evidence the presence of pores. The central void represents the lumen and the lateral voids within the flax cell walls represent the kink-bands. Several kink-bands can be observed in four different places (Fig 3. c-d, f-g). For example, the zoomed section on Figure 3h shows a close-up of the structure of the kink-bands and demonstrates that the cavities are crossing the fibre in all its thickness. The presence of cavities in kink-band regions has already been addressed by Melleli et al. (Melelli et al., 2021a) through SEM and AFM investigations and also by Wang et al. (Wang and Wang, 2005) on a large range of plant fibres including flax. Figures 3b and 3e represent a cross-section of the fibre without defects, only the lumens are detected, and no extra porosity is visible in these zones. In Figure 3 c-d, f-g irregular peripheral voids are detected revealing the kink-bands. The observed pores are globally long and narrow and are often curved towards the centre of the fibre. Depending on the cross section, the pores are encountered at different depths within the cell wall. For example, in Figure 3c, the porosity is spread across the cell wall depth. A simultaneous displacement along the axis of the fibres in longitudinal and transverse views, presented on Supplementary Information (Video 1), points out the correlation between the different observations, and especially the presence of pores within the kink-band volume when this latter can be longitudinally observed. These results confirm, in agreement with the hypothesis of Thygesen et al. (Thygesen et al., 2006) and Nyholm et al. (Nyholm et al., 2001), that kink-bands (also called dislocations) not only affect the structure of S1 but also S2-G layer.

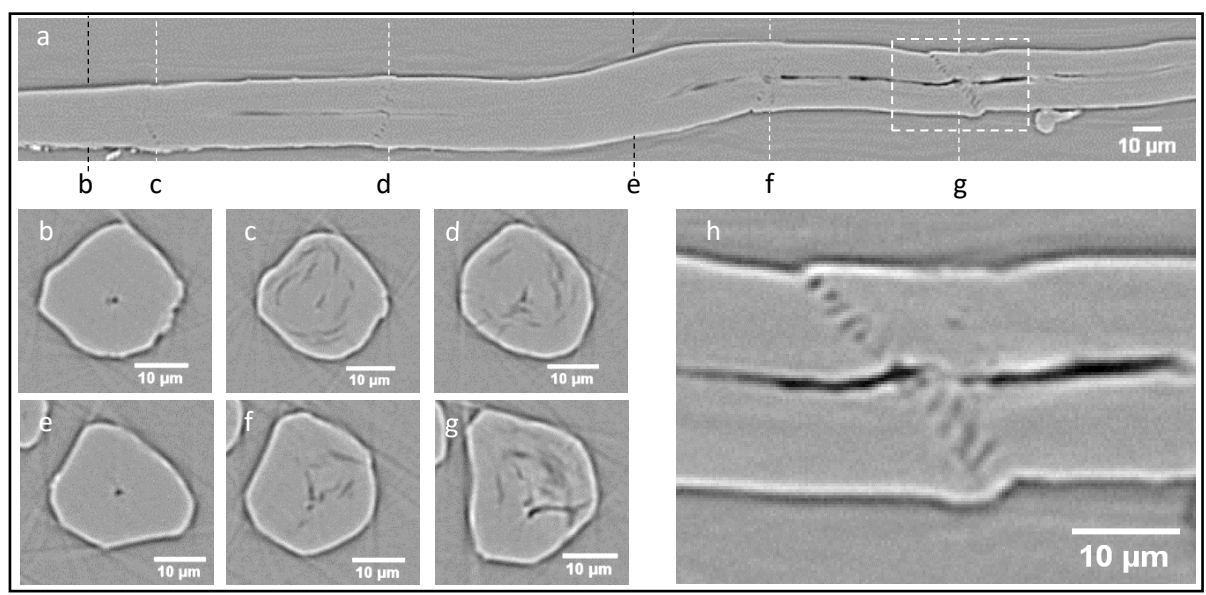

**Figure 3.** Example of a stretched fibre; visualization of kink bands in the fibre length through longitudinal and fibre cross section analysis

To visualize the morphology of the pores, 3D renderings of a drawn fibre bundle, constituted of four single fibres, are shown in Figure 4. The renderings clearly show the presence of kink-bands at the surface of the fibres (Fig. 4a) and Figure 4b shows the pores inside the fibres, including the lumens and the kink-bands in the cell walls. It is clearly visible that an important zone of pores is concentrated in the kink-band areas. Each enriched-porosity area in Figure 4b corresponding to a kink-band region; these regions are marked by the arrows in the surface rendering in Figure 4a. An area rich in porosity has been isolated in Figure 4c and clearly highlights the curved arrangement of pores.

To complement these results, Figure 5 shows a view of the same stretched fibre bundle, with a graphical presentation of the porosity content along the fibres. One can notice the low lumen volume, proving that the selected fibres are fully mature and cellulose filled (Goudenhooft et al., 2019); the lumen size value (~1%) is well correlated with literature values (Charlet et al., 2007; Richely et al., 2021). In kink-band regions, porosity ratio varies from 3 to more than 6% of the fibre cross section. To the best of our knowledge, these values are shown for the first time and confirm the strong structural damages in these zones, already visually observed with a range of techniques (Beaugrand et al., 2017; Rask et al., 2012; Wang and Wang, 2005).

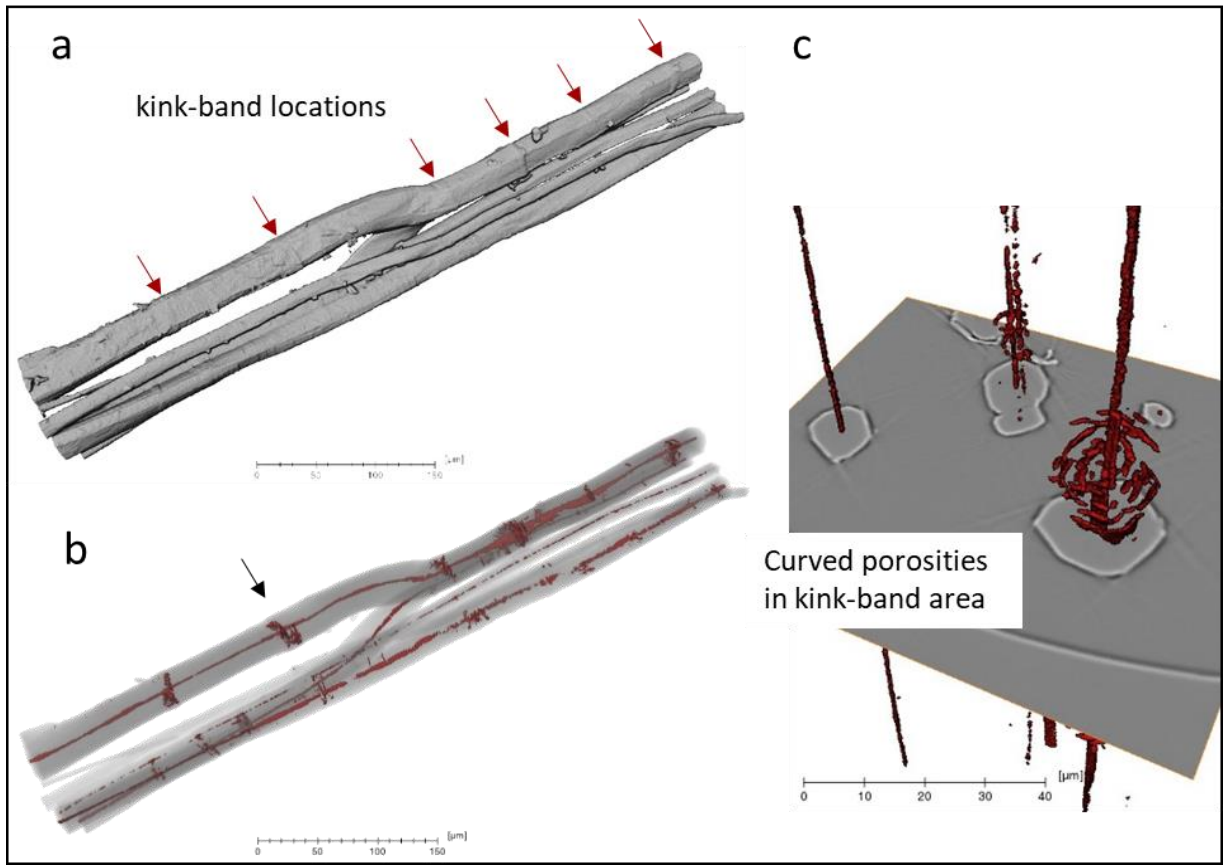

**Figure 4.** 3D renderings of drawn fibre bundles. (a) Surface of the fibres; red arrows indicate kink-bands. (b) Porosities are displayed in red inside the fibres. (c) Zoom on the kink-band located with the black arrow in panel b.

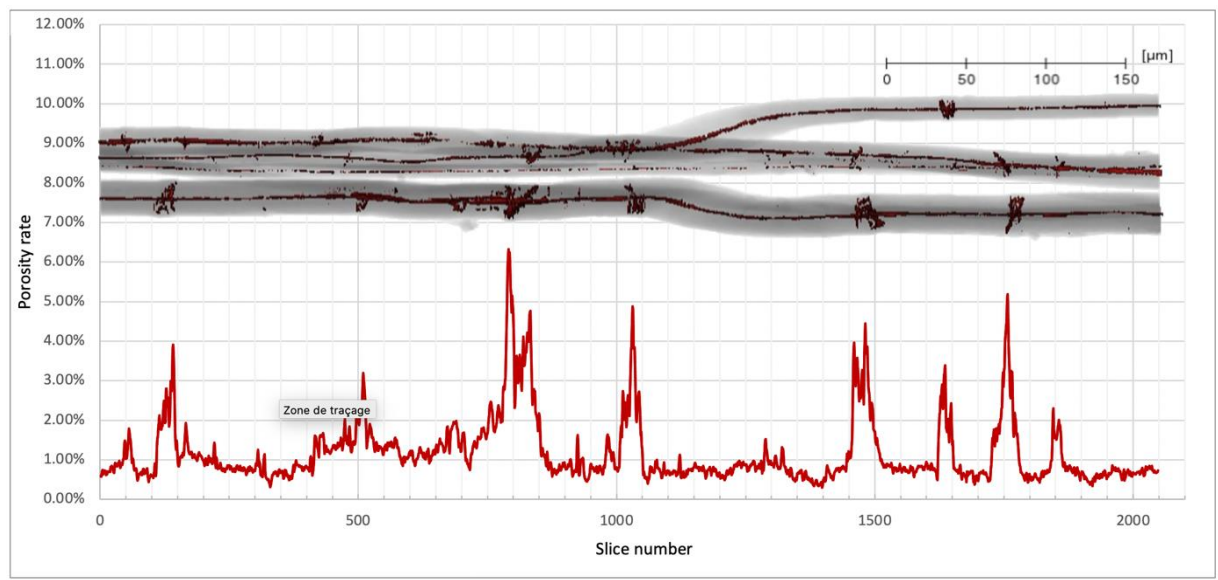

**Figure 5.** 3D renderings of a drawn fibre bundle and quantification of porosities along the fibres.

A local increase of porosity in the fibre S2-G layer is very likely at the origin of the degradation in mechanical properties of the fibres. The cavities, with sharp geometries, are expected to be places of stress concentration and favoured zones for crack initiation, therefore reducing the tensile strength of the fibre. The increase of porosity up to 6% should reduce the tensile modulus as it is the case in many porous materials whose properties are often decreasing exponentially as a function of the porosity rise (Ouagne et al., 2005).

3D X-ray tomography with adapted resolution, contrast and field of view is therefore a powerful tool to explore the kink-band structure and more specifically to visualise the distribution of pores in these regions.

### 3.3. Presence of kink-bands at the different stages of the fibre transformation process

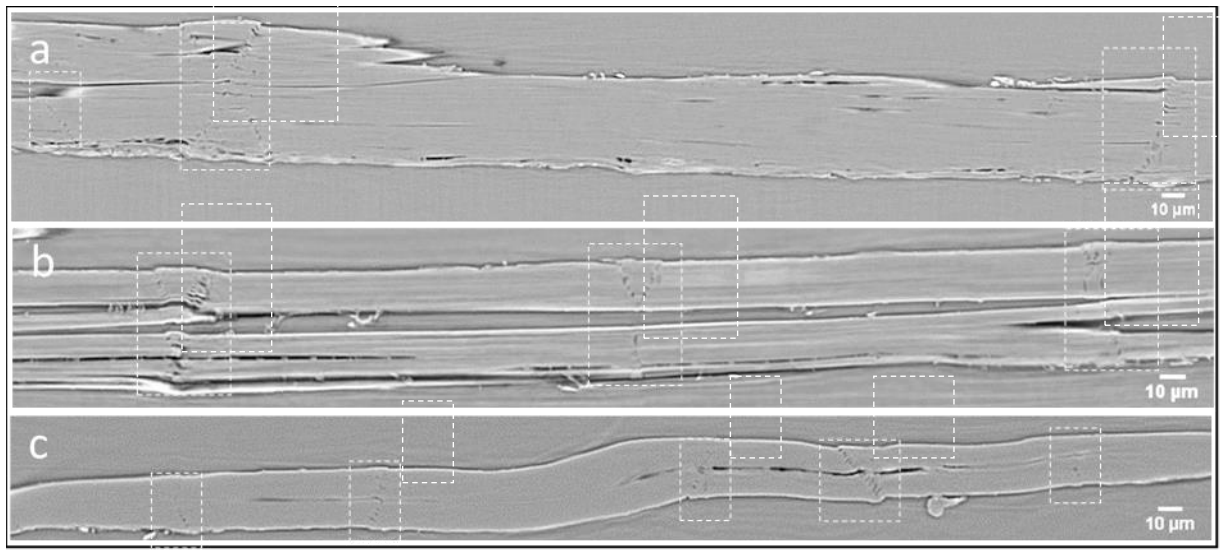

**Figure 6.** Visualization of kink bands (in dotted rectangles) in the fibre length through longitudinal and fibre cross section analysis. Scutched (a), hackled (b) and stretched (c) fibres were analysed.

In the previous section, pores inside the fibres have been detected and visualized on a drawn fibre corresponding to the last step of the process described in Fig. 1. As mentioned in the introduction part, the appearance or the multiplication of kink-bands may be due to mechanical loads applied to the stems and fibres during the fibre extraction and fibre preparation towards reinforcement manufacturing. It is interesting to study the impact of the prior process on the appearance and the morphology of the pores. Figure 6 represents single longitudinal tomographic slice images of elementary fibres or fibre bundles at different stages of the process: (a) after scutching, (b) hackling and (c) stretching. Kink-bands are detected at each step of the process with a number and an arrangement of kink-bands that are not drastically modified with the stage of the process; previous literature (Kozlova et al., 2022) demonstrated that the kink-band density was highly increased after the hackling step but in the present work, our objective is only to address the presence of kink-bands with a few images and not to finely quantify them. It can therefore be assumed, in agreement with SEM observations (as shown in Fig. 2), that kink-bands are either generated during fibre extraction (scutching) or were already present within the fibres before the fibre extraction step.

### 3.4. Investigation at the stem scale

As stated in the introduction part, the origin of kink-bands is not clearly determined. This work proposes to investigate if kink-bands are present in fibres within the stem, *in-planta*, (before fibre extraction). As received green stems and green stems submitted to severe bending, were investigated. Figure 7 shows longitudinal and radial cross sections of a bent green stem and of a dew retted stem. Pores are visible in the images, but these are inter-fibre pores which are due to possible lack of cohesion between the fibres within the stem (Fig. 7a, c, g). These pores were probably generated during bending of the stem as explained in the literature (Bos et al., 2002; Nodder, 1922). In the case of dew-retted stems, this is probably normal as consequent parts of the pectic cements were damaged during this phase. In Figure 7, no kink-bands can be observed in longitudinal and radial cross sections of dew-retted stems. This suggests that the kink-band defects are not generated if the fibres remain in the green or dew-retted stems even if the stems were submitted to severe bending. In Figure 7, green stems scans were analyzed, and the tomographic reconstruction does not show evidence of the presence of any kink-bands. Only full cell walls were observed in this case and inter pores between fibres in bundle Fig. 7g.

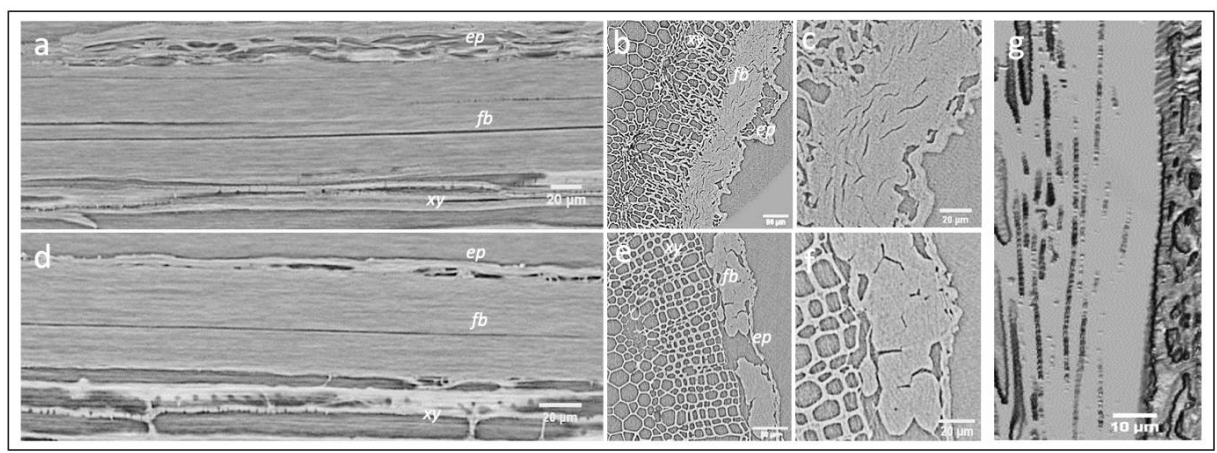

**Figure 7.** Comparison between bended green stem and dew-retted stem. Longitudinal (a) and transverse (b, c) sections of the bent green stem. Longitudinal (d) and transverse (e,f) sections of the dew-retted stem. Longitudinal view of a 3D reconstruction of the bent green stem (g). Fb: Fibre bundles; Xy: Xylem; Ep: Epidermis.

In addition, the supplementary material #2-4 (Videos 2, 3 and 4) presents the evolution of longitudinal and cross section views along green (Video 2), bent green (Video 3) and retted (Video 4) stems. In all these videos, it is never possible to observe porosity regions, especially in kink-band areas. Nevertheless, one can notice that fibre lumens are sometimes visible, proving again the interest of tomography for fibre internal structure analysis and also that with the available resolution, it would be possible to observe cavities inside kink-bands regions, if they existed. Our findings support the hypothesis of Kozlova et al. (Kozlova et al., 2022) regarding the fact that kink-bands appear during the mechanical extraction phase and not *in-planta*, even in case of severe bending of the stem. However, our results do not corroborate other studies, which suggest that kink-bands are developing in the plant stem before extraction, following environmental stress such as lodging induced by wind (Hendrickx, 2019; Thygesen et al., 2006).

### 3.5. Investigations of fibres extracted from green stem

As mentioned previously, no kink-band formation was observed *in-planta* either in green stem or after dew-retting process, suggesting that kink-bands are generated during the fibre extraction rather than by environmental conditions. Figure 8 shows longitudinal (a) and axial (b) cross-sections of green flax fibres manually extracted from the stem. Several defect areas can be observed along the fibre with irregular inner cavities and surface deformations revealing kink-band formation. Complementarily, these cavities are also addressed through Video 5. Un-retted fibres extracted from green stems show the presence of kink-bands on the contrary to fibres investigated within the stem. At the fibre scale, the presence and the number of kink-bands are not related to the level of retting as the defects were evidenced in both retted and un-retted fibres.

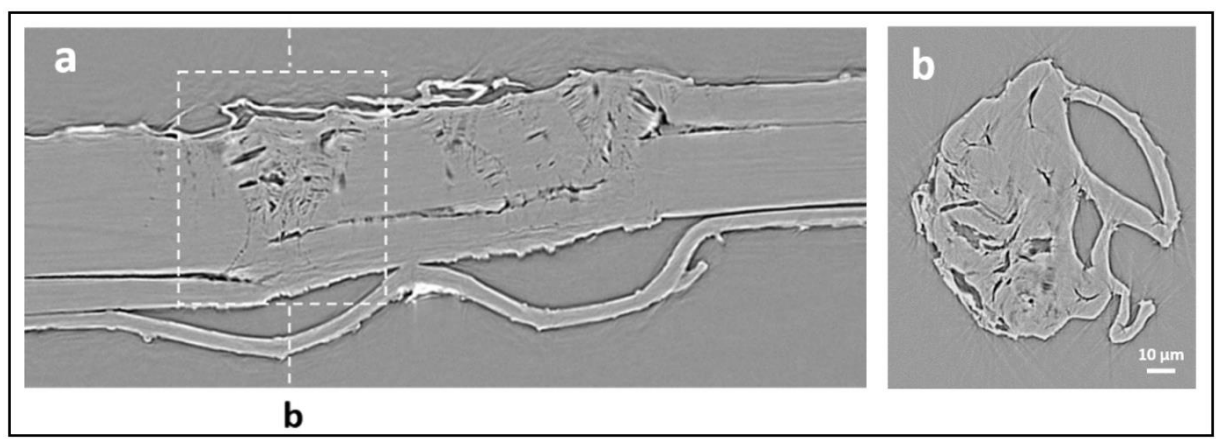

**Figure 8.** Visualization of defects in the fibre length through longitudinal (a) and axial (b) fibre cross-section images.

### 3.6. Preliminary analysis of the appearance of kink-bands

As mentioned in the introduction part, different hypotheses were proposed to explain the generation of defects. Some authors (Thygesen and Asgharipour, 2008) explain that the origin of the kink-band is due to external stresses during the plant growth. In Figure 7, no kink-band defects are visible at the surface of *in-planta* fibres, even after severe bending of the stems. This may be explained by the fact that the level of bending is not sufficient to cause kink-bands to form. In the case of fibres bound together within the stem, the level of bending deformation is lower than in the case of fibres separated from the stem. The bending does not reach a sufficiently small radius of curvature.

In the case of fibres extracted using scutching and hackling, the bending strains can be much higher and the radius of curvature reached can be much smaller. As the fibres are not bound together and bound to the woody core of the stem, the bending stiffness of the fibre is very low, and high strains may be reached in this deformation mode during scutching, hackling and drawing.

The appearance of kink-bands takes place only during the extraction of the fibres, when the fibre bundles are not anymore bound to the woody core of the stem. So, it appears that the hypothesis of kink-band formation *in-planta* during the drying/rehumidification cycles of the dew retting phase is not appropriate in our case. Kink-bands are visible in all the mechanically extracted fibres after scutching, hackling and drawing. During the scutching, the fibres are submitted to important loads leading to strong bending and longitudinal buckling of the fibre bundles. When the fibres are still bound together and to the woody core of the plant, the bending strains are not sufficient for creating significant kink-bands or cellulose macrofibrils misalignment. Indeed, the too large radii of curvature of the fibres and the longitudinal buckling is not a favoured mode of deformation, the fibres-woody core binding delaying this mode of deformation after other ones.

## 4. Conclusion

The presented work shows that kink-bands can be advantageously observed using synchrotron X-ray tomography. Kink-bands can be evidenced on elementary fibres and on technical fibres (fibre bundles) coming from retted and un-retted flax stems. Kink-bands can be observed as extra matter on the fibre surface as it is classically observed using tools such as SEM, but this defect can also be identified by the internal porosity contained within the cell walls at the location of the kink-bands. Kink-bands were identified on elementary fibres and on fibre bundles coming from different stages of the composite roving preparation value chain. No evidence of an increase in the number of kink-bands was observed during the succession of fibre extraction and preparation processes. Globally, the kink-bands are present on fibres coming after the scutching process for retted and un-retted fibres. Dew retting does not affect the presence or the global number of kink-bands. No kink-bands were observed on retted or un-retted fibres still contained within the flax stem even after severe bending deformation. This suggests that the kink-bands are not generated when the fibres are still within the stem and the hypothesis of kink-band appearance due to drying and re-humidification internal stresses is not confirmed by this work. On the contrary, this study shows that the kink-bands are generated during the fibre extraction processes such as scutching when fibres are separated from the woody core of the stem. In this case, the fibre bending stiffness is low and longitudinal buckling as well as severe bending takes place. When contained in the stem the longitudinal buckling does not take place and the bending is not severe enough (radius of curvature too high) to permit the generation of defects.

**Acknowledgements**

The authors wish to acknowledge the funding provided by the INTERREG IV Cross Channel programme through the FLOWER project (Grant Number 23). ANATOMIX is an Equipment of Excellence (EQUIPEX) funded by the *Investments for the Future* program of the French National Research Agency (ANR), project *NanoimagesX*, grant no. ANR-11-EQPX-0031.

Access to Anatomix was provided through SOLEIL beamtime proposal #20201291. The authors also wish to thank the Agence de l'eau Seine Normandie for supporting L. Pinsard's PhD.